\documentclass[]{emulateapj}

\newcommand{\degK}{\,{}^\circ\mbox{K}}
\newcommand{\grad}{^{\mbox{\small o}}}

\newcommand{\cm}{\mbox{\,cm}}

\newcommand{\kpc}{\mbox{\,kpc}}
\newcommand{\dyn}{\mbox{\,dyn}}

\newcommand{\Gyr}{\mbox{\,Gyr}}
\newcommand{\muG}{\,\mu\mbox{G}}
\newcommand{\kms}{\mbox{\,km s}^{-1}}

\newcommand{\eff}{\mathit{eff}}

\slugcomment{Accepted for publication in AJ}

\begin{document}

\title{Errors in kinematic distances and our image of the Milky Way
Galaxy.}

\author{Gilberto C. G\'omez}
\email{g.gomez@astrosmo.unam.mx}

\affil{Centro de Radioastronom\1a y Astrof\1sica - UNAM,
       Apartado Postal 3-72 (Xangari),
       Morelia, Mich. 58089, M\'exico}

\begin{abstract}

Errors in the kinematic distances, under the assumption of circular
gas orbits, were estimated by performing
synthetic observations of a model disk galaxy.
It was found that the error is $< 0.5 \kpc$ for most of the disk
when the measured rotation curve was used,
but larger if the real rotation curve is applied.
In both cases,
the error is significantly larger at the positions of the spiral
arms.
The error structure is such that, when kinematic distances are used
to develope a picture of the large scale density distribution,
the most significant
features of the numerical model are significantly distorted or
absent, while spurious structure appears.
By considering the full velocity field in the calculation of the
kinematic distances, most of the original density structures can be
recovered.

\end{abstract}



\section{INTRODUCTION}

Since the classic work by \citet{oor58}, there have been many
attempts to use the kinematic properties of the diffuse gas to
determine the large scale spiral structure of the Milky Way.
Very early in the study of the Galaxy,
it was determined that the orbits of the disk components of the
Galaxy were not very different from circular, with an orbital
frequency that decreases monotonically as a function of
galactocentric radius.
These facts allow the use of the kinematic distance method as a
first approximation to map the gaseous component of the galactic disk.
Two of the main strengths of this method (that it can be used
for a very large fraction of the Galaxy and that it can be
applied to the gaseous component of disk, which is notoriously
difficult to obtain a distance to) make it particularly useful for
this goal.
Nevertheless, it was soon realized that the deviations from circular
orbits, however small in absolute value, might have a strong impact
on how we see the Galaxy.

One of the first difficulties of the kinematic distance method
appeared in the determination of the rotation curve, namely,
in the fact that the circular rotation velocity measured for
positive galactic longitudes (northern Galaxy) did not match the one
measured for negative longitudes (southern Galaxy).
The simplest way to reconcile these observed rotation laws
is to take their average,
assuming that the differences generated by non-axisymmetric
structure will cancel out.
\citet{ker62} showed that this approximation leads to large north-south
asymmetries that, given their heliocentric nature, seemed unlikely.
It became clear that the complex
kinematic structure revealed in the diffuse gas surveys,
like the presence of gas at forbidden velocities, or the oscillations
in the rotation curve, was itself a consequence
of the spiral structure that was being sought.
Given the importance of the rotation curve for the understanding of
the galactic dynamics, in addition of the determination of kinematic
distances, a different approach to measuring the rotation curve was
needed.

A frequently used method to obtain the galactic rotation curve
involves measuring the full velocity field of discreet sources that
might share the motion of the diffuse gas (young objects like
\ion{H}{2} regions, for example), and then averaging the
so obtained azimuthal velocities.
This approach has lead to models of the rotation curve
\citep{bra93,mac05} that might more closely trace the real
mass distribution of the Galaxy, but introduce new
sources of error when used to obtain kinematic distances.
Nevertheless, generally accepted models of the spiral structure
\citep{geo76,tay93} have been obtained using this assumption.
\citep[For a nice review of the early work, see][]{ker69}

Another approach involves the
modelling of the non-circular motions of the gas, instead of forcing
the assumption of circular orbits.
In a recent paper, \citet{fos06} used an analytic approach for the
velocity field of the outer Galaxy.
In this work, a numerical model of the galactic disk, with full MHD,
is used to further explore the effects of
non-circular motions in the image one would obtain of the Galaxy
when we rely on the kinematic method for distances.
An observer is imagined inside the numerical model,
which is assumed similar to the Milky Way, and the analysis that this
observer would perform is reproduced.
In \S \ref{simulation_sec}, a brief description of the
numerical simulation is presented;
in \S \ref{observations_sec} the selection of the observer's position
is described, and how the measurement of the rotation
curve was emulated;
\S \ref{errors_sec} presents an analysis of the errors in the kinematic
distances and how they affect the image the observer generates of
his/her home galaxy;
finally, \S \ref{discusion_sec} summarizes the results.


\section{THE SIMULATION}
\label{simulation_sec}

The numerical model used here is described elsewhere
\citep{mar04a, mar04b, yan05}, and only an overview is presented here.

The initial setup consisted of a gaseous disk with an
exponential density profile in the radial direction, with a scale
length of $4 \kpc$.
The density at the position of the Sun (at $8 \kpc$ from the
galactic center) was taken to be $1.11 \cm^{-3}$.
The equation of state for the gas was isothermal with
$T = 10^4 \degK$.
This disk was threaded by an azimuthal magnetic field,
with strength given by the relation,

\begin{equation}
p_B = p_M \frac{n}{n+n_c},
\label{pM_eq}
\end{equation}

\noindent
where $p_B$ is the magnetic pressure, $n$ is the gas density,
$p_M = 1.43 \times 10^{-12} \dyn \cm^{-2}$,
and $n_c = 0.04 \cm^{-3}$.
Equation (\ref{pM_eq}) yields $B = 5.89 \muG$ for $r=R_\odot$.
This intensity, and the  magnetic field geometry, were set
only as initial conditions, and were allowed to evolve in time
according to the ideal MHD equations.

The gas initially follows circular orbits, with a velocity given by
the equilibrium between the background gravitational potential,
the thermal and magnetic pressures, and the magnetic tension:

\begin{equation}
\frac{v_\phi^2}{r}
  = \frac{1}{\rho} \frac{\partial (p_T + p_B)}{\partial r}
  + \frac{\partial \Phi}{\partial r}
  + \frac{2 p_B}{\rho},
\end{equation}

\noindent
where $v_\phi$ is the azimuthal velocity,
$\rho = m_{\eff} n$ is the gas mass density,
$m_\eff = 1.27 m_H$ is the mean particle mass,
$p_T$ is the thermal pressure,
and $\Phi$ is the gravitational
potential described by \citet[ model 2]{deh98}.

The equilibrium was then perturbed by the two-armed
spiral potential described in \citet{pic03}.
The simulation was performed in the perturbation reference
frame, which rotates with an angular speed $\Omega_P = 20 \kms
\kpc^{-1}$ \citep{mar04a, mar04b}.
It is worth mentioning that the perturbing potential was calculated
as a superposition of oblate spheroidals, and so it does not have
the usual sinusoidal profile.
Also, the parameters that describe the
perturbation (total mass in the arms, pitch angle, pattern speed,
etc.) were constrained by those authors so that the pattern is
self-consistent in the stellar orbits sense.

The MHD equations were solved using a version of the ZEUS code
\citep{sto92a, sto92b},
a finite difference, time explicit, operator split, eulerian code
for ideal magnetohydrodynamics.
The numerical domain consisted in a two-dimensional grid in
cylindrical geometry, with $500^2$ points.
The numerical domain extended from $1 \kpc$ through $15 \kpc$ in
radius, and spanned a full circle in azimuthal angle.
The boundary conditions were reflecting in the radial direction.

\begin{figure}
\plotone{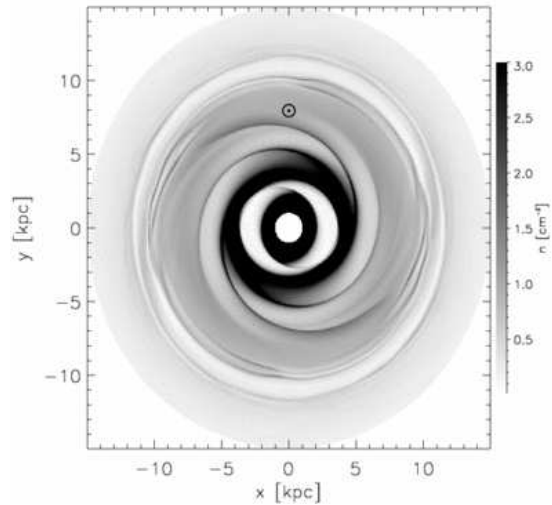}
\caption{
Density distribution of the
simulation after $1 \Gyr$ of evolution.
The grayscale shows the gas density in units of $\cm^{-3}$.
It is noticeable that, although the perturbation has only two arms,
the gas response shows four arms.
The Sun symbol shows the position chosen for the imaginary observer.
}
\label{initial_fig}
\end{figure}

Figure \ref{initial_fig} shows the simulation after $1 \Gyr$ of
evolution.
The most important characteristics of the simulation at this stage
are:
1) although the perturbation consists of two spiral arms, the
gas forms four arms (two pairs with pitch angles of $9\grad$ and
$13\grad$ each, as opposed to the perturbation with a pitch angle of
$15.5\grad$);
2) a high density ring is formed at $r \approx 4 \kpc$; and,
3) a low density ring is formed near corotation, at
$r \approx 11 \kpc$.
Again, details of the simulation and the physical phenomena related
to these structures are discussed elsewhere.
The corotation low density ring was found
(using a different background potential) by Martos \& Y\'a\~nez
(private communication).
A study of the neccesary conditions for the formation of such ring,
its physics and consequences for our Galaxy is presented in
Martos (2006, in preparation).

\citet{gom04a, gom04b} also performed large scale simulations of the
Galaxy.
Since their numerical model was three-dimensional, they were able to
study some phenomena (like the vertical motions associated to the
hydraulic jump behavior of the gas near the spiral arms) that could
have an impact in the dynamics of the gas near the midplane.
Nevertheless, the focus of their model was to study those phenomena,
and an emulation of the Milky Way Galaxy was not a priority.
Specifically, their three-dimensional numerical grid, a necessity in
their work, restricted the spatial resolution achivable in the
midplane.
This, together with the low value for $\Omega_P$ used,
did not allow the formation of four spiral arms as a response
to a two-arm spiral potential
(in order to obtain four gaseous arms, their model included a
four-arm potential).
In the present work, the model was restricted to the galactic plane
so that sufficient resolution can be reached.


\begin{figure}
\plotone{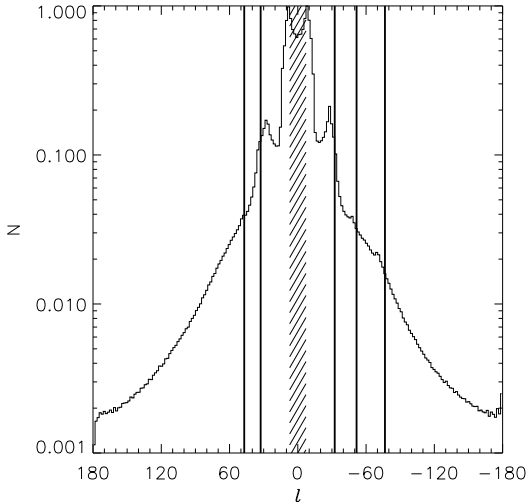}
\caption{
Column density, in arbitrary units,
vs. galactic longitude in the simulation.
By moving the imaginary observer around a circle of radius
$R_\odot = 8\kpc$, the local maxima can be matched to the observed
directions tangent to the spiral arms (thick vertical lines).
The dashed area at $|l| < 7 \grad$ shows the simulation inner radial
boundary.
}
\label{tangents_fig}
\end{figure}

\section{THE SYNTHETIC OBSERVATIONS}
\label{observations_sec}

Local maxima in column density ($N$) vs. galactic longitude ($l$)
plots for the diffuse gas are
usually interpreted as the directions at which the line-of-sight is
tangent to spiral arms.
The $N$ vs. $l$ distribution an imaginary observer would see if
placed inside this model galaxy is presented in
Figure \ref{tangents_fig}.
By moving the observer around the solar circle, at $8 \kpc$ in the
numerical model,
the number and positions of the local maxima can be fitted to the
observed values for the diffuse gas.
In this work, the chosen directions were those tangent to the locus
of the spiral arms proposed by \citet{tay93}, namely,
$l = -76.5\grad, -51.6\grad, -32.4\grad, 32.7\grad$ and $47.1 \grad$.
It was found that by choosing the position shown in Figure
\ref{initial_fig} for the imaginary observer all but one
of the column density local maxima in the model
fall within $3\grad$ of these quoted directions.
If the tangent directions quoted by \citet{dri01} are adopted,
namely $l \approx -80\grad, \pm50\grad$ and $\pm30\grad$,
all but one of the tangent directions yield an even better fit.
(The difference between the ill-fitting tangent in the model,
at $l \approx -72\grad$, and the quoted direction is in fact smaller
than the width of the feature observed in $240 \,\mu\mbox{m}$.
See, for example, \citealt{dri00} and \citealt{dri01}.)


\subsection{The rotation curve}
\label{rotcurve_sec}

Once a position for the observer is chosen,
the next step toward calculating the kinematic distances is to
adopt a rotation curve for the simulated galaxy.
For the inner galaxy ($r < R_\odot$),
the standard procedure consists in searching for the terminal
velocity of the gas, i.e., the maximum line-of-sight component
of the velocity (minimum, for negative longitudes).
If one assumes that the gas orbits are circular, the terminal
velocity arises from the point at which the line-of-sight is
tangent to the orbit, and so, the galactocentric radius of
the emitting gas is known.
Under this assumption, the circular rotation curve for the galaxy is
given by:

\begin{equation}
v_c(r) = v_t(l) + v_\odot \sin(l),
\label{circvel_eq}
\end{equation}

\noindent
where $v_c$ is the circular velocity, $v_t(l)$ is the terminal
velocity for a given galactic longitude,
$v_\odot$ is the velocity at the solar circle,
and $r = R_\odot \sin(l)$ is the galactocentric radius of the
tangent point.

At this point, a choice between two options for the value of the
circular velocity at the solar circle has to be made.
One option is to take the circular velocity consistent with the
background gravitational potential ($v_\odot = 220 \kms$).
This option has the disadvantage that the gas in the evolved
simulation will stream by the observer
[although this is not necessarily wrong, since the presence of
gas at forbidden velocities in the $l-v$ diagram is well known
\citep{lin67, bli91}].
Nevertheless, it was decided in this work to take a second option,
which is to take the value for $v_\odot$ ($ =225\kms$)
given by the azimuthal velocity of the
gas in the evolved simulation at the position assigned to the
imaginary observer,
since this choice would more closely mimic the procedure used to
determine the Local Standard of Rest from galactic sources
\citep{bin98}.
There will still be streaming gas, but this will
happen in the radial direction only.
Such radially streaming gas has been reported by \citet{bra93}.

\begin{figure}
\plotone{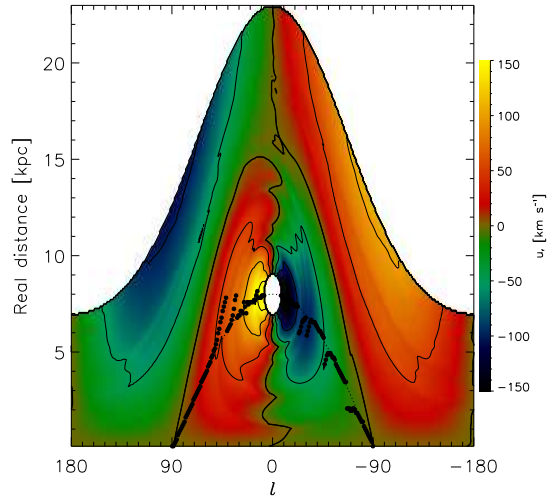}
\caption{
Line-of-sight component of the velocity field as a function of
galactic longitude and (real) distance to the observer, with
contours every $50 \kms$ (the thicker contour marks the $u_r =
0\kms$ level).
The dotted line shows the locus of the tangent
points, while the circles show the positions at which the
terminal velocity is reached.
}
\label{lon_dist_fig}
\end{figure}

Figure \ref{lon_dist_fig} shows the line-of-sight component of the
velocity field.
The Figure also shows the actual positions at which the terminal
velocity is reached for a given $l$.
Although the distance between those terminal-velocity points and the
tangent points is typically small, the non-circular motions and
spiral shocks generate kpc-scale deviations and discontinuities in
the terminal-velocity locus.
Since those deviations happen at the positions of the spiral arms,
they will generate larger errors at the vicinity of the arms,
and will strongly affect the observer's view of the spiral structure
of the model galaxy.

For the outer galaxy ($r > R_\odot$), the usual procedure to
determine the rotation curve involves looking
for sources with independently known distance and measuring their
line-of-sight velocity \citep[ for example]{bra93}.
This procedure was simulated by assuming that the observer finds
such a source at each point of the numerical grid outside the solar
circle.
It is assumed that the distances to such sources are less
reliable the farther they are from the observer.
So, the circular velocity for the outer galaxy was taken to be,

\begin{equation}
v_c(r) = \frac{r}{R_\odot}
  \left( \sum\limits_\phi \frac{w_\phi v_{los}}{\sin(l)}
  + v_\odot \right),
\label{circvel_out_eq}
\end{equation}

\noindent
where the weights $w_\phi$ decrease linearly with the distance to
the observer, and the summation is performed, at a given radius,
over the azimuthal points excluding those within $7 \grad$ of
the galactic center and anti-center directions.

\begin{figure*}
\plottwo{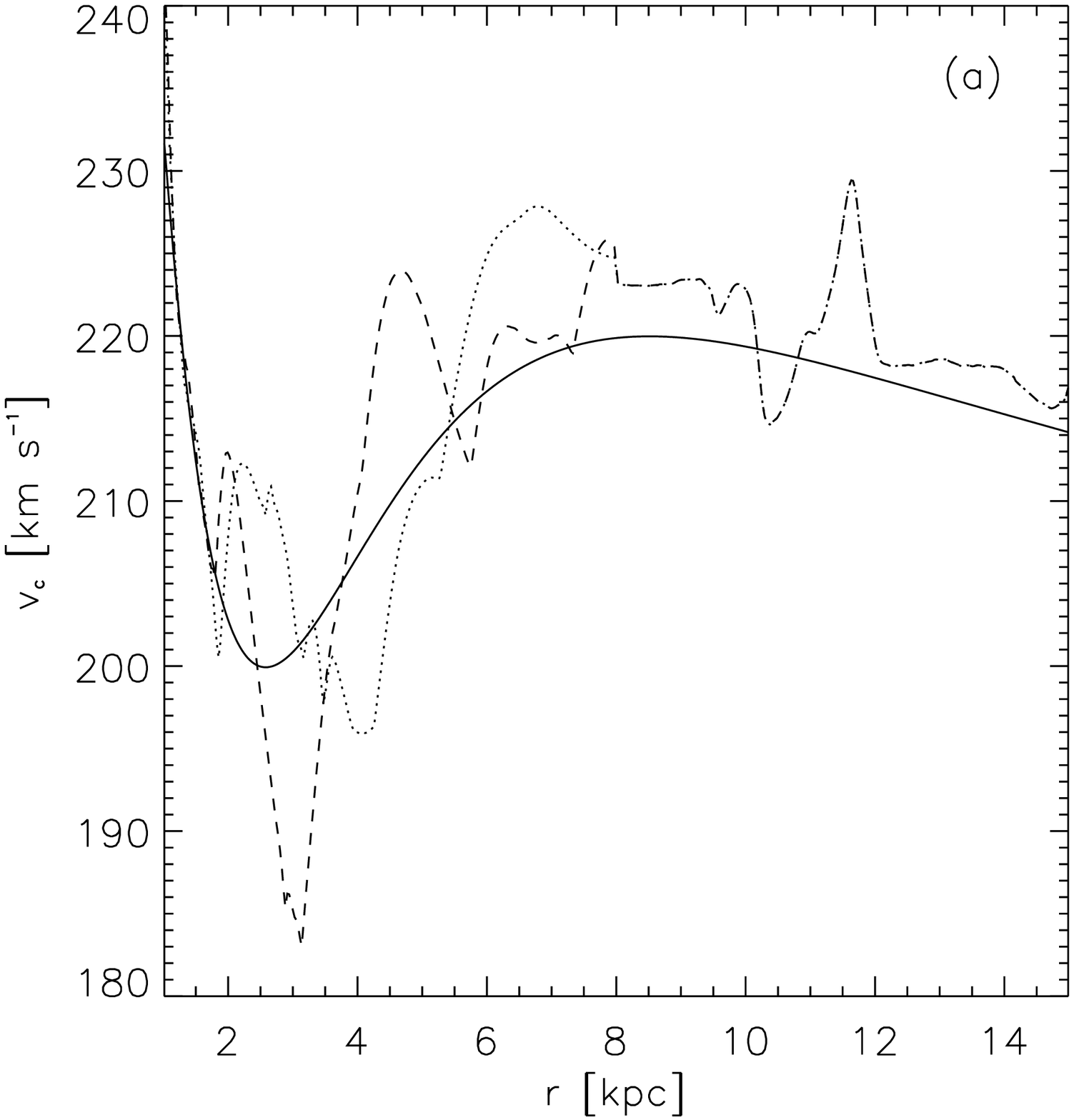}
        {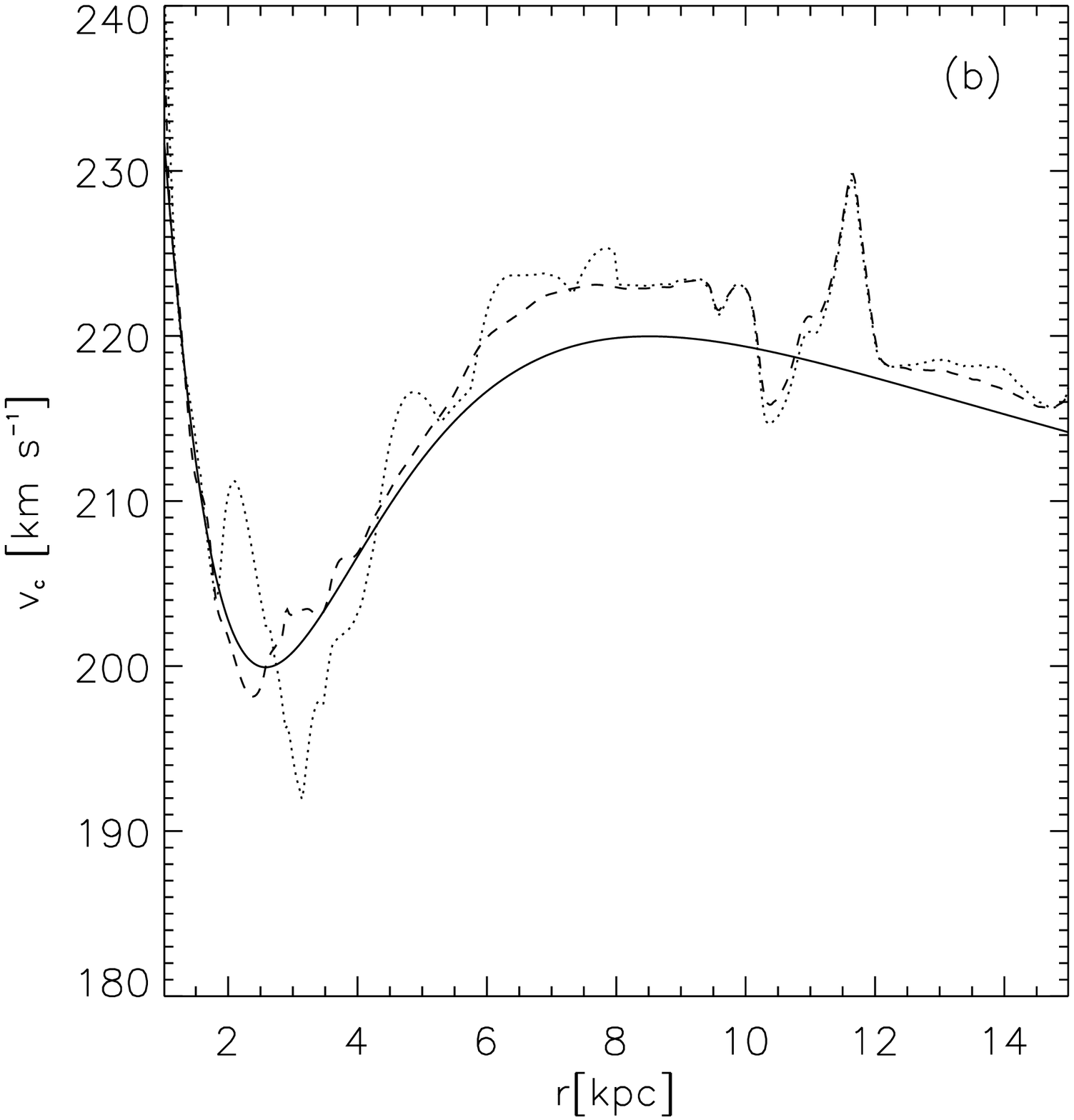}
\caption{
a) The rotation curve given by the background potential (solid line),
compared with the measured rotation curve.
For the inner galaxy, both the rotation measured for positive
(dotted line) and negative longitudes (dashed line) are presented.
The rotation curve corresponding to the outer galaxy is also shown
(dashed-dotted line).
b) The rotation curve given by the background potential (solid line),
compared with the average of the north and south rotation curves
(dotted line) and the mean azimuthal velocity of the gas in the
simulation (dashed line).
Notice that both the mean velocity and the mean rotation curve are
above the background rotation curve for most of the radial domain.
}
\label{rot_curve_fig}
\end{figure*}

Figure \ref{rot_curve_fig}(a) shows the so obtained rotation curves,
together with the rotation consistent with the background
gravitational potential.
The northern rotation curve is lower than the southern rotation at
$\sim 3.5\kpc < r < \sim 5.5\kpc$, while the opposite is true
up to $r = R_\odot$.
This behavior is similar to the rotation curves reported by
\citet{bli91}, when scaled for $R_\odot = 8\kpc$.

In order to try to recuperate the true (background) rotation,
that should more closely trace the large scale mass distribution,
the average both northern and southern rotation curves was taken.
The result is compared with the background rotation in Figure
\ref{rot_curve_fig}(b).
Although the result is smoother and closer to the rotation
consistent with the background potential, it is still systematically
higher \citep[in agreement with the results reported by][]{sin78}.
Another approach is to take the full velocity field and
average the azimuthal velocity of the gas
\citep{bra93}.
The result, also shown in Figure \ref{rot_curve_fig}(b), is much
closer to the background rotation, but it is still systematically
larger.


\section{ERRORS IN THE KINEMATIC DISTANCE}
\label{errors_sec}

After adopting a rotation curve, and assuming that the gas
follows circular orbits, the errors in the measured kinematic
distances can be estimated by comparing the measured with the real
distance in the simulation.

In order to resolve the distance ambiguity for the inner galaxy, the
usual procedure is to bracket the distance close
enough as to place the object of interest on either side of
the tangent point by looking at the galactic lattitude extension
of the source \citep{fis03}, or using observed intermediate
absorption features \citep{wat03, sew04}.
For this investigation, I decided to cheat: I looked up on which
side of the tangent point the gas parcel fell on, and chose the
measured distance accordingly.

\begin{figure*}
\plottwo{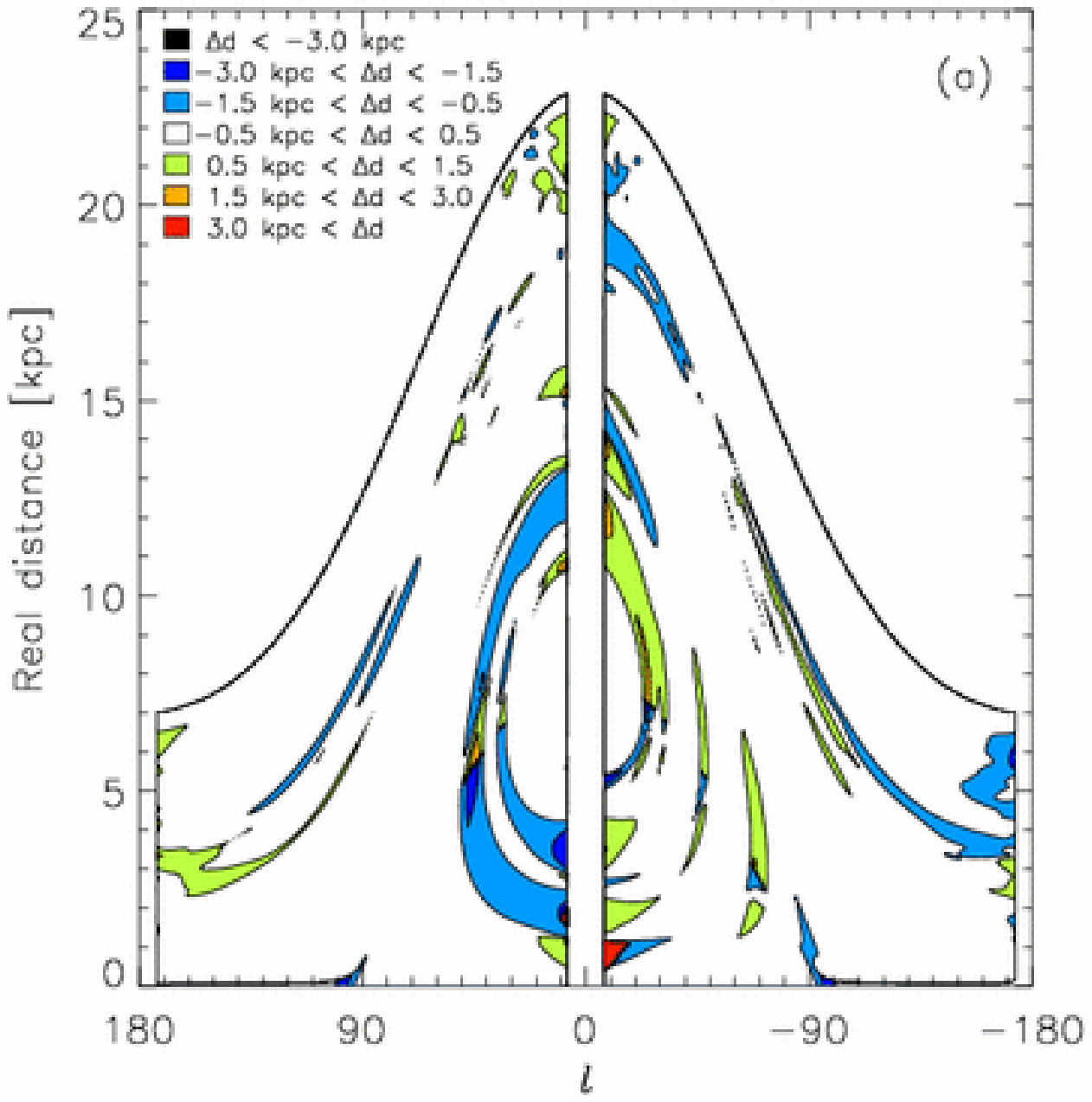}
        {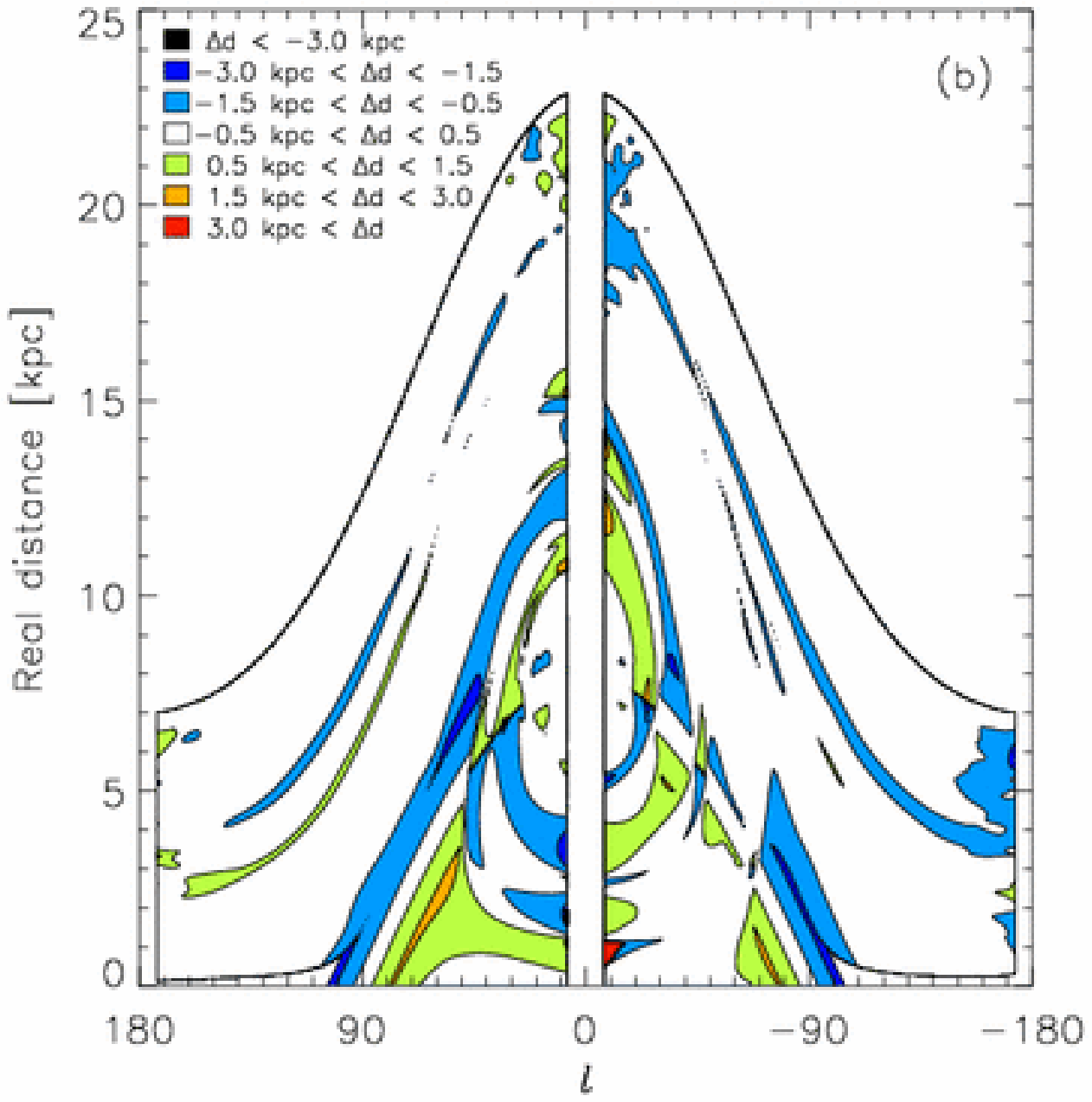}
\caption{
Error in the measured kinematic distance ($\Delta d$) obtained under
the assumption of circular orbits following (a) the measured rotation
curve, and (b) the rotation given by the background potential.
Sections of $7 \grad$ around the galactic longitudes $l=0\grad$ and
$180\grad$ were excluded.
Although the error in most of the galactic disk is of the order of
$0.5 \kpc$, it is significantly larger at the positions of the
spiral arms.
The sharp edges at the position of the tangent points are a
consequence of the fact that the terminal velocities do not happen
at those points.
The errors in measured kinematic distances are larger when the real
(background) rotation curve is used.
}
\label{lon_err_fig}
\end{figure*}

Figure \ref{lon_err_fig}(a) shows the error in measured distance
with respect to the real distance in the model.
Recalling Figure \ref{rot_curve_fig}, the observer would determine
different rotation curves for the northern and southern sides of the
galactic center.
Accordingly, in determining the kinematic distance for Figure
\ref{lon_err_fig}(a), the rotation curve used was that of the
corresponding side of the galactic center.
It is noticeable that although the errors are of the order of
$0.5 \kpc$ in most of the galactic disk,
they are significantly larger at the positions of the spiral arms
\citep[as hinted by][]{gom04b}.
This fact has a special impact in studies of the spiral structure
of the Galaxy that rely on kinematic distances, since it distorts
the image the observer would generate (see \S\ref{distorted_sec}).

There is another significant feature in Figure \ref{lon_err_fig}.
Although the terminal velocity does not really arises from the
tangent point, the circular orbits assumption assigns gas observed
near terminal velocity to that point.
This fact generates a feature in the errors that
corresponds to the locus of the tangent points.
Again, the error is significant at the position of the
spiral arms and would generate large errors in the determination of
distances to objects that trace the spiral structure.

The assumptions of circular orbits and different rotation
curves for positive and negative longitudes are, of course,
inconsistent.
One solution is to fit a single
rotation curve to both sides of the Galaxy.
In order to test this method, the average of both rotation curves
was taken and the equivalent of Figure \ref{lon_err_fig}(a) was
calculated.
The result was that the magnitude of the error in the kinematic
distances was approximately the same, but the area with
error $> 0.5 \kpc$ spanned a larger fraction of the disk.

Suppose now that
the imaginary observer somehow manages to obtain the large scale
distribution of stellar mass in the model galaxy.
This would allow the derivation of the real rotation curve from
the background axisymmetric potential.
If the observer now uses that real rotation to determine kinematic
distances, even larger distance errors would be
obtained, specially for the inner galaxy, as shown in
Figure \ref{lon_err_fig}(b).
This counter-intuitive fact arises because, at this point of the
simulation, the gas has already adopted orbits that are not only
influenced by the background potential, but also the large scale
magnetic field (likely different from the field in the initial
conditions) and the torques and resonances generated by the spiral
perturbation. 
Although the real rotation curve is consistent with the most important
determinant of the gas rotation velocity (the background mass
distribution), it does not include other influences in that
velocity, while the ``wrong'' rotation curve determined from gaseous
terminal velocities more closely reflects the real motion of the gas
(recall Fig. \ref{rot_curve_fig}, where the measured rotation curve
is systematically above the true rotation).

Although intrinsically inconsistent, the two different measured
rotation curves were used in the remaining of this investigation
since that procedure leads to smaller distance errors.
The results presented in the following section are even more
notorious if the average or the real rotation curves are used.


\subsection{The Galaxy Distorted.}
\label{distorted_sec}

\begin{figure}[b]
\plotone{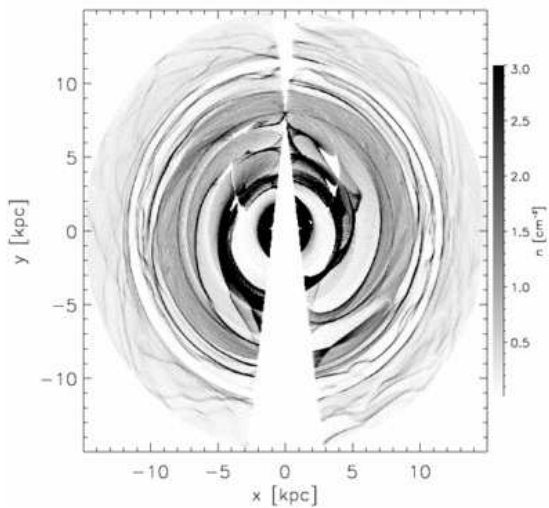}
\caption{
Re-mapping of the gas distribution resulting from the kinematic
distances using the measured rotation curves in Figure
\ref{rot_curve_fig}(a) and assuming circular gas orbits.
Notice the regions near the tangent points and the corotation circle,
where little or no gas is mapped to.
}
\label{deforma_fig}
\end{figure}

Consider now that the imaginary observer
tries to study the spiral structure of the galaxy he/she lives in.
The procedure would consist of translating the longitude-velocity
data obtained from a diffuse gas survey, for example,
into a spatial distribution using the kinematic
distances that result from the assumption of circular orbits that
follow the measured rotation curve.\footnote{
In order to diminish spurious interpolation effects, each gas parcel
was spread using a 2D gaussian weight function into a $3\times3$
grid-cell region around the position corresponding to that parcel's
galactic longitude and measured distance.}
The resulting map is shown in Figure \ref{deforma_fig}.
Notice that the features described for Figure \ref{initial_fig}
(namely the four spiral arms, the $4 \kpc$ high density ring, and the
corotation low density ring)
all but disappear, while new fictitious features, like the structure
in the outer Galaxy, are formed as a consequence of the
oscillations in the outer rotation curve.
Also significant in this Figure are the regions where little or no
gas is assigned by the mapping, namely the bands near the corotation
circle and the quasi-triangular regions near the tangent point locus.
(This nearly empty regions are significantly larger when the
background or the mean rotation curves were used to determine the
distance to the observed gas parcel.)

The imaginary observer would likely conclude that his/her home
galaxy has 2 ill-defined spiral arms.
If a logarithmic spiral model were forced,
an $\sim 11\grad$ pitch angle and a density contrast much stronger 
than that in the numerical model would be found.

Another possibility for determining the distance to a gas parcel
consists in comparing the line-of-sight velocity of the parcel with
the predicted velocity obtained from some model for the
galactic structure.
For the numerical model described in \S \ref{simulation_sec}, given
a galactic longitude, Figure \ref{lon_dist_fig} is searched
for the required velocity, and the corresponding distance is
read out.\footnote{
A simple C language program that provides a distance given a
galactic longitude and line-of-sight velocity value and uncertainty
is available at
{\tt http://www.astrosmo.unam.mx/\~{}g.gomez/publica/}.
In that program, the resulting distance is given as a range, instead
of a central value and uncertainty, since the velocity-distance
mapping makes the distance probability distribution neither uniform
in the range, nor peaked around a central value.}
Although the same procedure to solve the ambiguity with
respect to the tangent point was used, the non-circular motions
introduce new distance ambiguities for certain longitude-velocity
values (up to 11, although 3 is a more typical number).
When these ambiguities appear, they happen close to each other,
making their resolution difficult.
So, when reconstructing the map of the galaxy, the gas density was
equally split among these positions.

\begin{figure}
\plotone{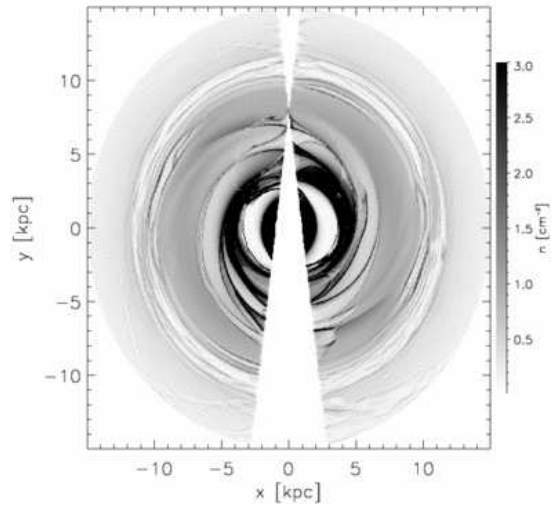}
\caption{
Same as Figure \ref{deforma_fig}, but using the full velocity field
to recover the density distribution.
Most of the characteristics of Figure \ref{initial_fig} are recovered,
although some spurious structure appears due to the new distance
ambiguities introduced by the non-circular motions.
}
\label{deforma_rec_fig}
\end{figure}

The result is shown in Figure \ref{deforma_rec_fig}.
The new distance ambiguities still introduce spurious structure,
like the splitting of the spiral arms.
Nevertheless, the number and position of the arms, the
structure around the corotation radius and the lack of features in the
outer galaxy are recovered.
The imaginary observer would likely conclude that his/her home
galaxy has 4 arms with $9\grad$ and $12.5\grad$ pitch angles,
although he/she would also find non-existing bridges and spurs.
On the other hand, it should be considered that the imaginary
observer does not see thermal nor turbulent line broadening.
When these are considered, some of the new ambiguities will be
swallowed into a distance range, effectively dissapearing.
Therefore, some of the spurious structures will blend with real
structures.
So, the observer might get an image of the model galaxy closer to
reality than Figure \ref{deforma_rec_fig} suggests.


\section{SUMMARY AND DISCUSSION}
\label{discusion_sec}

The effect of the circular orbits
assumption on our idea of the large scale structure of the Galaxy
was explored.
Since these errors might be quite large at the position of the
spiral arms, the study of the spiral structure of the Galaxy and
objects associated with it is particularly affected.
By simulating the way an imaginary observer inside the model galaxy
might try to infer the structure of the gaseous disk, it was found
that the circular orbits assumption destroys the spiral structure
and creates spurious features in the measured distribution.

The method of kinematic distances is a powerful one since it allows
measurement of distances to diffuse sources and it is easily
applicable to a large fraction of the galactic disk.
Even if the measured rotation curve includes deviations that do
not reflect the true large scale mass distribution,
Figure \ref{lon_err_fig}(a) shows that the errors in the
distance are, in fact not very large for most of the galactic disk;
in fact, the distance errors that arise from using the true
rotation curve are larger.
In both cases, however, the errors are quite large at the positions
of the spiral arms.
If we want to use this distance method for objects associated with
the spiral structure, we need to consider non-circular motions
(as has been succesfully shown by \citealp{fos06} for a set of
\ion{H}{2} regions and SNR).

One possibility to achieve this is to try to determine the full
velocity field of the galactic disk.
But direct measure of distances to the diffuse gas component is quite
difficult (therefore the strength of the kinematic distance method).
So, we need to use discreet objects and assume that they share their
velocity with the diffuse component (\citealp[ for example]{bra93};
\citealp[see also discussion in][]{min73}).
Yet another difficulty arises when 
tangential velocities and distances are required beyond the
solar neighborhood.

Another approach at determining the full velocity field is to 
model it.
Recently, \citet{fos06} used an analytical model of the density and
velocity fields of the diffuse gas,
with parameters for the model fitted to \ion{H}{1} observations 
of the outer Galaxy.
Despite the fact that their density and velocity models are not
consistent in the hydrodynamics sense, and that the model do not
include the dynamical effects of magnetic fields, they were able to add
features of the Galaxy that are currently difficult to incorporate to
numerical models, like the disk's warp or the rolling motions
associated with the spiral arms.
Further numerical studies should allow the development of a more
realistic analytical model.

Instead of an analytic model, a numerical model was used in the
present work to obtain density and velocity fields.
Since the focus is in large scale velocity structures, an
eulerian code provides a good approach.
Also, since the galactic magnetic field has been proved to be an
important component of the total ISM pressure \citep{bou90},
its effect in the gas dynamics is likely to be important;
therefore, a full MHD simulation was required.
The large scale forcing is also trascendant;
since the azimuthal shape of the spiral perturbation appears to have
an influence on the gaseous response \citep{fra02},
the usual sinusoidal perturbation was deemed too simplistic and
a self-consistent model for the perturbing arms was chosen.
At the present time,
the galactic warp and the vertical motions associated with the spiral
arms \citep{gom04a,gom04b} could not be considered at the necessary
resolution.

In this work, it has been shown that it is possible to recover most
of the gaseous structure of the galactic disk using kinematic
distances, as long as the full velocity field is considered.
Nevertheless, applying these results to the Milky Way is a whole new
issue, since obtaining the full velocity field is not trivial.
For the procedure used here,
how close the numerical simulation is to the real Galaxy
remains the weak point of this approach.
The computation cost of a realistic enough simulation is still too
high to allow a parameter fitting analysis.
So, the remaining question is
if the velocity field that results from the simulation
yields a determination of the distance to a given object,
or only an estimation of the distance error.
The answer to that question is left to the reader's criterion.


\acknowledgements

This author wishes to thank
J. Ballesteros-Paredes,
E. V\'azquez-Semadeni,
C. Watson,
J. Franco,
L. Loinard,
S. Kurtz, and
an anonymous referee
for their encouragement and useful comments during the preparation
of this manuscript.


{}


\end{document}